\documentclass[aps,prl,amsmath,amssymb,reprint,superscriptaddress]{revtex4-1}
\usepackage{graphicx}
\usepackage{amsmath}
\usepackage{dcolumn}
\usepackage{bm}
\usepackage{bbold}
\usepackage{tabularx}
\usepackage{xcolor}

\bibstyle{apsrev4-1}

\begin{document}
\title{Observation of Low Temperature Magneto-Mechanic Effects in Crystalline Resonant Phonon Cavities}

\author{Maxim Goryachev}
\email{maxim.goryachev@uwa.edu.au}
\affiliation{ARC Centre of Excellence for Engineered Quantum Systems, University of Western Australia, 35 Stirling Highway, Crawley WA 6009, Australia}

\author{Serge Galliou}
\affiliation{Department of Time and Frequency, FEMTO-ST Institute, ENSMM, 26 Chemin de l'\'{E}pitaphe, 25000, Besan\c{c}on, France}

\author{Michael E. Tobar}
\affiliation{ARC Centre of Excellence for Engineered Quantum Systems, University of Western Australia, 35 Stirling Highway, Crawley WA 6009, Australia}

\begin{abstract}

We observe magnetic effects in ultra-high quality factor crystalline quartz Bulk Acoustic Wave resonators at milli-Kelvin temperature. The study reveals existence of hysteresis loops, jumps and memory effects of acoustical resonance frequencies. These loops arise as a response to the external magnetic field and span over few Hertz range for modes with linewidths of about $25$mHz, which constitute a frequency shift of order 60 linewidths. The effects are broadband but get stronger towards higher frequencies where both nonlinear effects and losses are limited by two level systems. This suggests that the observed effects are due to ferromagnet-like phase of a spin ensemble coupled to mechanical modes. The observed coupling between mechanical and spin degrees of freedom in the ultra low loss regime brings new possibilities for the emerging class of quantum hybrid systems.

\end{abstract}

\date{\today}
\maketitle

\section{Introduction}

Magnetic effects in mechanical resonators have become an interesting subject of research with many potential applications. In particular the combination of magnetic properties of solids and high quality factor mechanical resonators has set a new direction in the sensing technology\cite{Gross:2016aa,Bian:2016aa,Weber:2012aa,Vanner18}, such as cantilever magnetometry\cite{Weber:2012aa,PhysRevLett.111.067202,Montinaro2014,Mehlin2015}. Moreover, careful engineering of coupling between spin and mechanical degrees of freedom\cite{Rabl:2009aa} opens new possibilities in hybrid quantum systems\cite{Kurizki:2015aa} where one can combine advantages of optomechanics and Quantum Electrodynamics with two level systems\cite{Xia:2014aa,PhysRevA.95.022327} or magnons\cite{Zhang:2016aa,PhysRevLett.1.241}. Such research leads to better understanding of magnetic impurities and their influence on mechanical properties of solids that, in its turn, creates fundamental knowledge for a new generation of ultra-high purity materials with reduced susceptibility to environmental changes, ageing and other unwanted effects. 

To observe such effects in a high purity acoustic system, the most important property is to attain very high quality factors of mechanical resonances. The resulting very narrow linewidths become a very sensitive tool in the discovery of low concentrations of impurities whose effects would not be seen otherwise. Among all mechanical systems, quartz Bulk Acoustic Wave (BAW) resonators cooled to cryogenic temperatures provide the narrowest possible linewidths for many overtone (OT) resonances in the $5-700$~MHz frequency range\cite{Goryachev1,quartzPRL,ScRep}, and more recently similar performance at GHz frequencies\cite{Renninger:2018vl}, allowing strong coupling to photons and superconducting qubits\cite{Enzian:19,Kharel2018,Chu:2018sf}. In this work, we study response of these mechanical cavities to external magnetic fields for temperatures near $20$~mK.

\section{Bulk Acoustic Wave Resonator Under Magnetic Field}

\begin{figure}[h!]
     \begin{center}
            \includegraphics[width=0.25\textwidth]{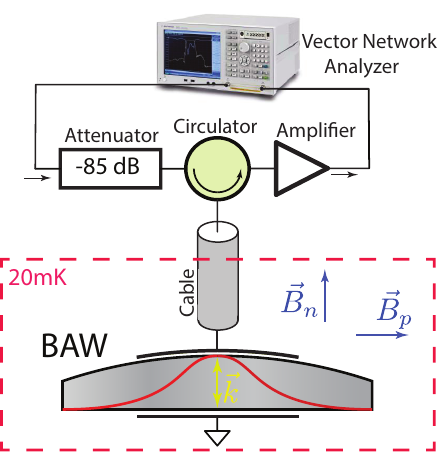}
            \end{center}
    \caption{Experimental setup for measuring frequency response of BAW cavity with acoustic modes standing along $\vec{k}$ as a function of external magnetic field that is parallel ($\vec{B}_p$) or normal ($\vec{B}_n$) to the resonator plate.}%
   \label{setupSIG}
\end{figure}

For these measurements we choose an SC-cut\cite{Kusters:2014mn} plano-convex electrodeless (BVA type\cite{1537081}) resonator enclosed in a copper vacuum chamber. While the plano-convex structure of the resonator plate works as a potential trap for the acoustic phonons minimising losses due to supporting mechanisms\cite{Tiersten:1976hz,Stevens:1986aa}, the electrodeless (electrodes are deposited on the supporting structure) technology helps to exclude all losses associated with deposition of other materials. 
It has been discovered previously that the longitudinal {\color{black}(A-type)} thickness mode exhibits the superior quality factor compared to shear modes reaching very high overtone numbers. For this type of mode that is used in this work, the mechanical displacement is mostly collinear with acoustical wave vector that is normal to the plate surface. {\color{black} On the other hand, for the shear ({\color{black} B- and C-type}) modes used later in the work, the displacement lies mostly in the resonator plane.}

In order to investigate response of the acoustic modes to external magnetic field, the quartz BAW cavity is enclosed into a block of Oxygen Free Copper attached to the $20$~mK stage of a dilution refrigerator. Because the crystal is enclosed in an individual metallic package inside a copper block, any effects of spurious electric fields, caused by the magnet or other external sources will be shielded. The structure is assembled in such a way that the device under test is placed in the middle of a $7$~T superconducting magnet. The resonator plate could be oriented either parallel or normal to the external magnetic field that are further referred as $\vec{B}_p$ and $\vec{B}_n$ respectively (Fig.~\ref{setupSIG}). 
The acoustic mode frequency (more precisely, frequency deviation from fixed values) is used as a convenient observable of the induced magnetic effects, as it is the most sensitive parameter of a resonant system. 

Acoustic resonances are excited and detected using the piezoelectric effects with golden electrodes deposited on the structure supporting the cavity plate. RF fields are supplied via low loss microwave cables split with cryogenic DC blocks. Since the investigated BAW resonator is a single port device, all possible measurements are made in signal reflection where one uses an impedance or vector network analyser to probe the frequency response of the device via the single port. Although, in  the case of ultra high-$Q$ acoustic cavities, internal attenuation of typical network analysers is not enough to achieve the linear (small signal) regime. 
Indeed, it has been observed previously that for a number of ultra-high Quality factor overtones, one needs to greatly reduce the incident power to achieve the small-signal regime. In this case, the feeding and detection lines are split with a circulator that allows to introduce additional attenuation and amplification in the two port measurements as shown in Fig.~\ref{setupSIG}. {\color{black}At low frequencies} ($<1$~GHz), circulators become lossy at low temperatures due to ferrite properties, its operation is limited to the room temperature. Moreover, comparatively narrow operation widths of these devices limits their operation to a couple of overtones near its central frequency. 

To characterise the acoustic resonance in the magnetic fields, we employ the 37th overtone of the longitudinal mode of the plate resonator with the resonance frequency of $116.1642$~MHz. For the small signal characterisation, the incident signal was reduced and distributed by a room temperature attenuator and circulator as depicted in Fig.~\ref{setupSIG}. Together with additional low noise amplification, this setup allows to achieve the incident power levels of about $-135$~dBm, or tenths of a femto-Watt, at the BAW crystal port. This value is estimated based on the specified losses of the connecting cryogenic superconducting cables. The given value of incident power (-135 dBm) refers to the input of the BAW crystal. We take into account losses of cryogenic (superconducting) connecting cables and DC blocks. Such low amounts of power are possible due to very narrow bandwidths (long sweeping rates) of the measurements. Indeed, a system response for a particular value of the magnetic field is taken with the lowest available bandwidth of 1Hz totalling about 26 minutes of sweep time. Such long measurement times are required to avoid ringing effects. The measurement setup is frequency locked to a hydrogen maser for ultimate frequency stability allowing us to resolve sub milli-Hertz frequency shifts at very long averaging times.

\subsection{Parallel Orientation}

\begin{figure}[h!]
     \begin{center}
            \includegraphics[width=0.45\textwidth]{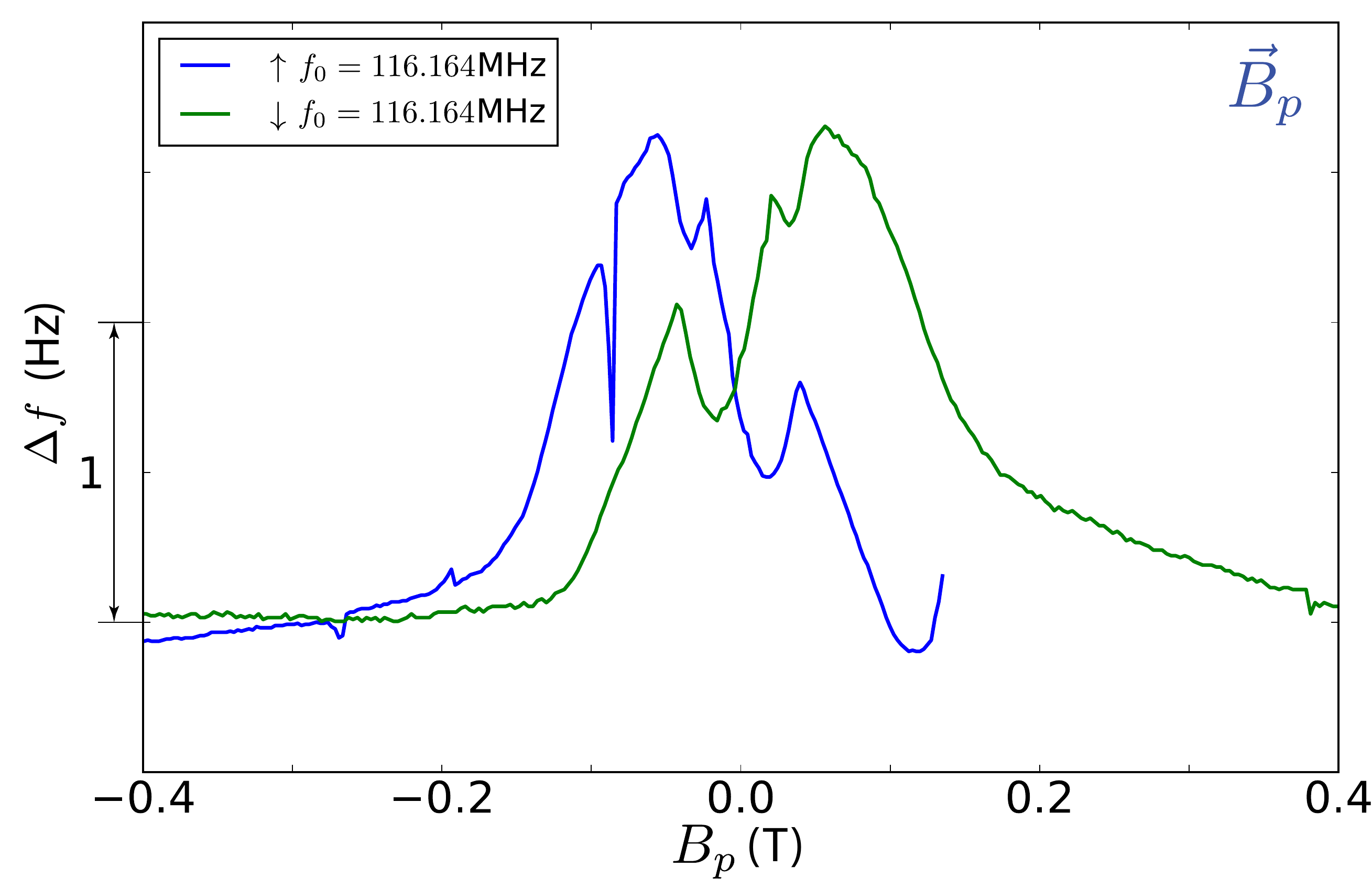}
            \end{center}
    \caption{Frequency deviation as a function of external magnetic field parallel to the crystal plate for sweeping magnetic field up and down for the 37th OT of the longitudinal mode.}%
   \label{sweepS}
\end{figure}

The response of the resonant frequency of the 37th OT to the magnetic field oriented in parallel to the resonator plate is shown in Fig.~\ref{sweepS}. The experiment is repeated by sweeping the magnetic field from negative fields to positive and vice a versa. The results show strong hysteresis of the response together with the memory effect. Two results are almost symmetric around the zero field. The frequency deviation gets saturated at relative large values of the field. Such behaviour including memory and hysteresis is typical for ferromagnetic system\cite{Jiles:1986aa,Prokopenko:2013aa,Farr:2015aa}.

\subsection{Normal Orientation}

\begin{figure}[h!]
     \begin{center}
            \includegraphics[width=0.45\textwidth]{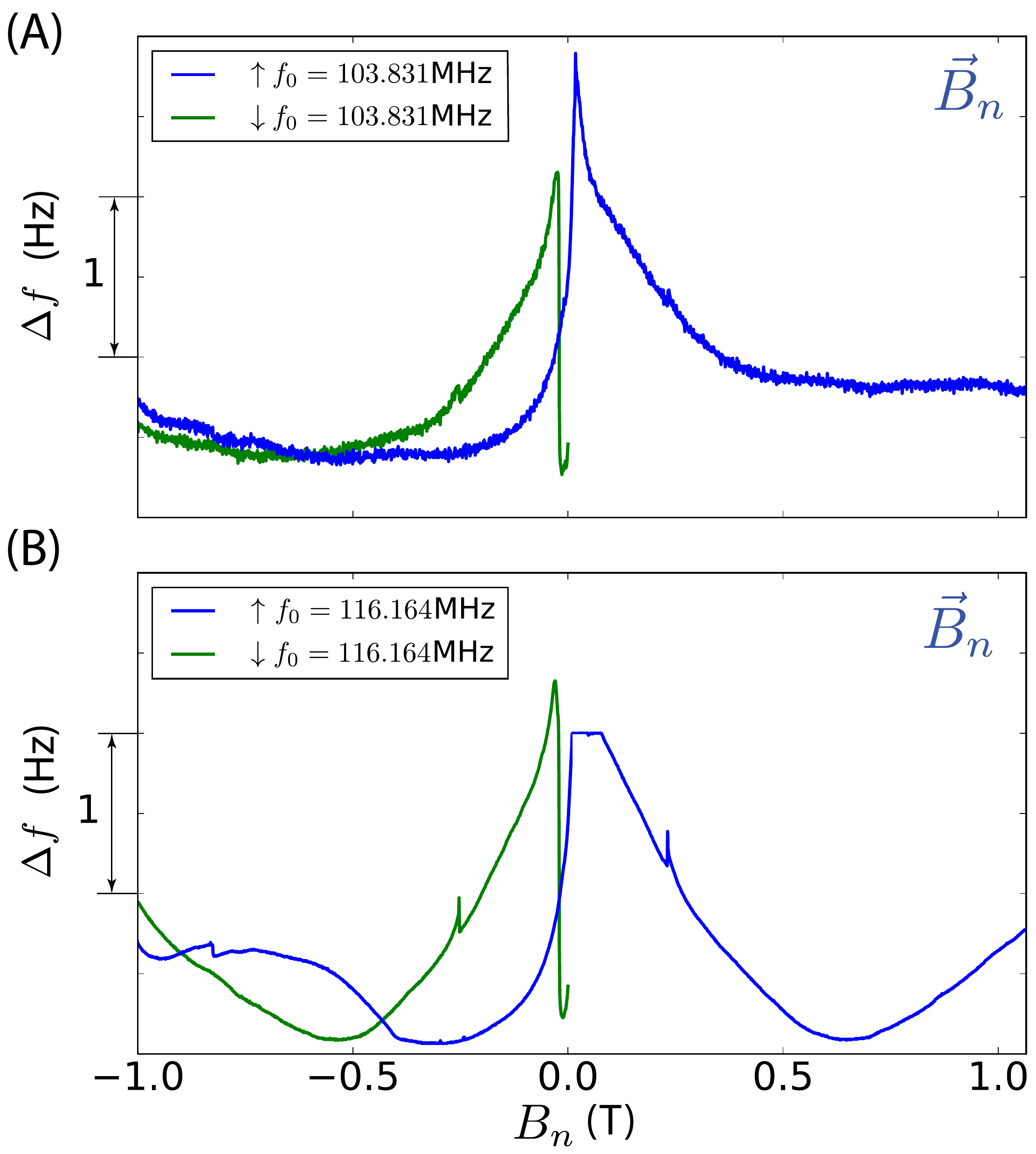}
            \end{center}
    \caption{Frequency deviation as a function of external magnetic field normal to the crystal plate for sweeping magnetic field up and down for the 33st (A) and 37th OTs (B) of the longitudinal mode.}%
   \label{sweepR}
\end{figure}
\begin{figure*}[h!]
     \begin{center}
            \includegraphics[width=1.0\textwidth]{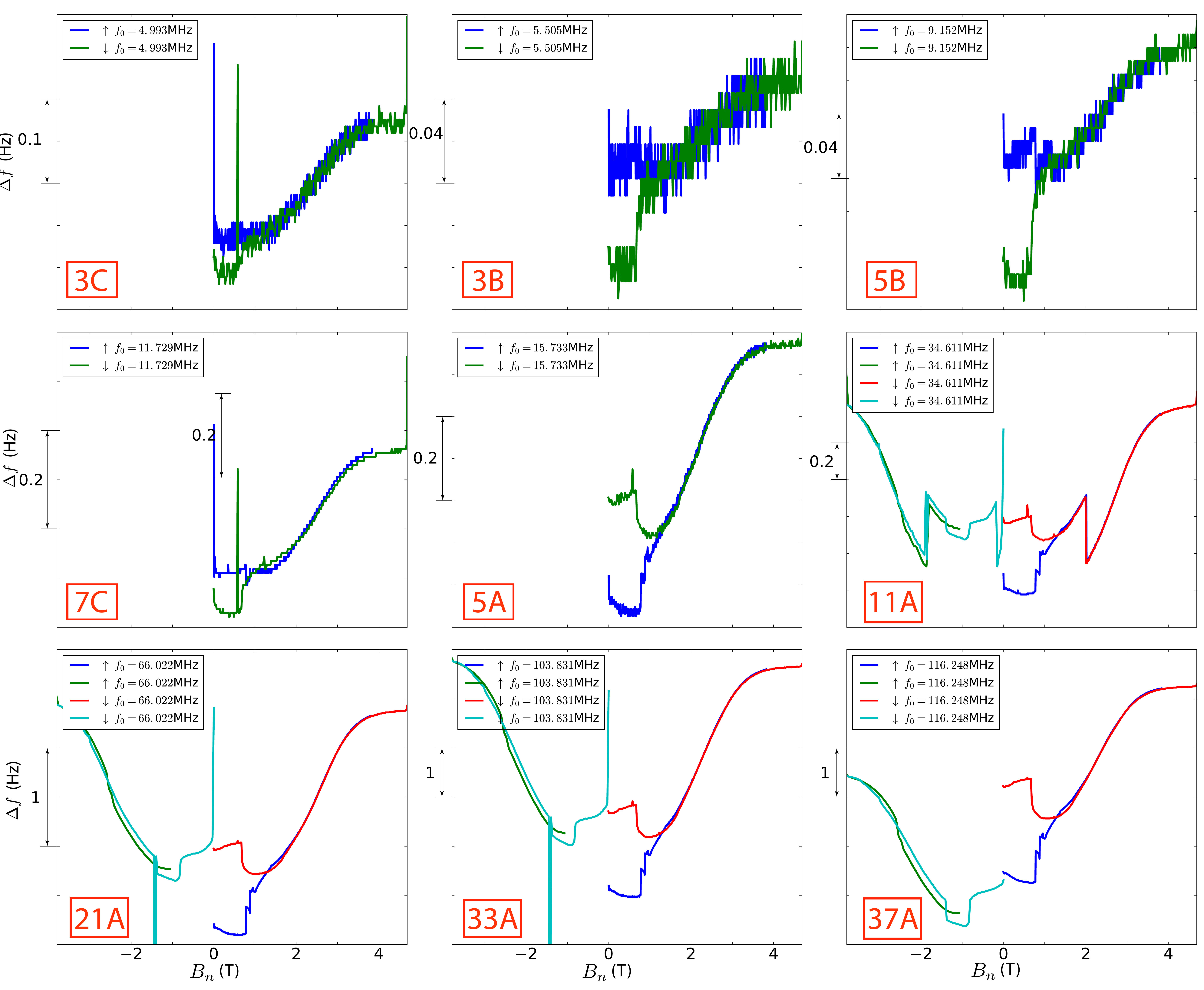}
            \end{center}
    \caption{Frequency deviation as a function of external magnetic field normal to the crystal plate in the strong signal regime. Indices in boxes denote OT numbers and types of vibration: A - longitudinal, B - fast shear, C - slow shear. }
   \label{sweepV}
\end{figure*}

Similar type of behavior is observed for the 33rd and 37th OTs in the normal orientation of the magnetic field as shown in Fig.~\ref{sweepR}. {\color{black}In these results, the field was swept down (green $\downarrow$ curve) and then up (blue $\uparrow$ curve).} For this field orientation the observed hysteresis {\color{black}width in terms of the applied magnetic field is considerably larger. The corresponding frequency deviation is of the same order.} The observed difference with the parallel orientation of the field suggest anisotropy of the magnetic system responsible for the mechanical frequency shifts. 
{\color{black}It is worth noting that two measurements ($\downarrow$ and $\uparrow$) in Fig.~\ref{sweepR} (A) and (B) have different frequencies at zero field as well as they are not symmetric under the rotation around zero field, thus, measurement prehistory, which suggests strong memory effects.} Despite the fact that the observed frequency shifts are on the order of few Hertz, given extremely narrow linewidths in studied BAW resonators ($25$~mHz for the 37th OT at 116.164MHz \cite{quartzPRL}), the observed effects appear to be very significant constituting about 60 linewidths. This makes the BAW cavities unique for study of such weak magneto-acoustic effects.

\section{Strong Signal Measurements}

In order to investigate the dependence of the magnetic hysteresis on absolute values of the frequency, the quartz resonator is probed directly with the Network Vector Analyser. Such simplified setup avoids circulators that allows to measure overtones of acoustic vibration in a very broad frequency range. As a result of impossibility to decrease the Network Analyser output signal strength below a certain limit, the device incident power is relatively strong on the order of $-45$~dBm or tens of nanoWatt. Such level of incident power results in observable nonlinearity of the Duffing type. Results of frequency response to magnetic field $\vec{B}_n$ sweeps for various resonant modes are shown in Fig.~\ref{sweepV}. This figure shows that the spread of the hysteresis loop increases with increasing frequency from tens of milli-Hertz for less than 10MHz modes to a couple of Hertz for high frequency modes. This fact also implies a broadband nature of the effect which is different to the linear interaction of impurity ion ensembles with photonic modes leading to narrow band avoided level crossings\cite{Goryachev:2013aa}. It is also observed that all modes demonstrate saturation at high magnetic fields. Finally, the response of 33rd and 37th OTs of the longitudinal mode in the strong signal regime is quite different to that in the weak signal setup shown in Fig.~\ref{sweepR} suggesting an important role of nonlinearity in the interaction between magnetic and acoustic subsystems. The observed hysteresis and memory effects have strong temperature dependence. They are specifically milli-Kelvin phenomena that vanishes at temperatures above 1K, so that no considerable frequency shift has been observed at 4K.  

\section{Discussion}

The response of quartz BAW resonators to the magnetic field at room temperature have been observed previously\cite{Brendel:1994aa}. It was attributed to ferromagnetic behaviour of supporting springs holding the resonator plate. The same mechanism cannot be responsible for the magnetic effects observed in this study for three reasons. Firstly, all magnetic effects are specific to temperatures below $1$K: no observable frequency shifts are visible at 4K where as clamping mechanism impact is observed even at room temperature. Secondly, the resonator used in the current work is built using a different clamping technology that avoids magnetic elements. Thirdly, the magnetic field effects are accompanied with strong nonlinear response to the excitation signals that should have no effect on supporting mechanism. Finally, increase in the OT number usually leads to improvement of the phonon trapping\cite{Goryachev:2014aa} that means that a larger fraction of acoustic energy is focused in the centre of the resonator place. This effect typically leads to increase in $Q$ factor with frequency due to uncoupling of acoustic vibration from support structure. This fact also implies that the supporting mechanism has less effect on resonance frequency for higher OTs, that contradicts the observed results (Fig.~\ref{sweepV}). 

Coupling between mechanical and spin degrees of freedom may be accounted for several effects such as magnetostriction, Einstein-de Haas, Wiedemann and related effects. Although all of these phenomena require ferromagnetic materials, the class of materials quartz does not belong to. The evidence strongly suggests that the magnetic effects observed in this work originates in crystal impurities themselves. Traces of magnetic impurities have been observed in pure synthetic quartz crystals using whispering gallery microwave modes\cite{Goryachev:2013aa}, and has been shown to be the mechanism for the largest magnetic response of the crystal material. Furthermore, the temperature dependence of mechanical Quality factors of quartz resonators showed a relaxation phenomenon due to the presence of Na$^+$, Al$^{3+}$ ions substituting some Silicon ions in the lattice\cite{Mason1965, Fraser1968,Martin:1988aa,Kats1962,Halliburton1985,Stevels1962}. These and other impurities in quartz have been a subject of extensive studies in literature\cite{Saha1979,Poignon:1996aa}. Moreover, two level systems were found to be responsible for strong nonlinearities in acoustic resonators made of unswept (non purified) quartz\cite{quartzJAP}, jump chaotic response of such devices at low temperatures\cite{Goryachev:2014ad}, and extra losses at higher frequencies\cite{quartzPRL}. Additionally, interaction of high frequency acoustic phonons with magnetic spins has been studied within the research field of paramagnetic acoustic resonance\cite{BOLEF:1966aa, Jacobsen:1959aa,apr2}.
 
 Although, the ferromagnetic-like behaviour was not observed in quartz studies with Whispering Gallery Mode (WGM) resonators\cite{Goryachev:2013aa}, there are a good reasons why this is the case. Firstly, concentration of defects and impurities depends on the crystal manufacturer, and the crystals used for WGM resonators and BAW devices came from different sources. In particular, crystal treatment plays a crucial role for impurity behaviour. For example, it has been experimentally observed that annealing of sapphire crystals can convert Fe$^{2+}$ ions into Fe$^{3+}$ which caused a number of significant new effects unobserved in "as-grown" crystals\cite{Creedon:2010aa}. Similarly, quartz crystals used for BAW resonators undergo the so-called sweeping procedure, whereas the quartz used in the WGM resonators had no such treatment. During the sweeping procedure quartz samples are heated to a few hundreds degrees and high voltage is applied to remove certain cations\cite{Martin:1988aa,Gualtieri1989SweepingQC}. 
Secondly, the BAW devices are highly sensitive, exhibiting tens of milli-Hertz acoustic linewidths, which is many orders of magnitude narrower than that of the dielectric linewidths of WGM resonators. While the former approach detected effects on the Hz scale, the WGM modes in quartz\cite{Goryachev:2013aa} have typical linewidths of tens of kHz leaving any effect of the order of 1Hz, like ones observed in this work, very hard to observe. 
 
In summary, the magnetic effects observed in this work have only become apparent due to the phonon trapped acoustic modes reaching extremely high quality factors of a few billion. Indeed, with other technologies providing quality factors of less than one million or linewidths of a few hundreds of Hertz, such hysteresis loops observed in this work could not have been seen. These high acoustic $Q$-factors makes the phonon trapped BAW resonators a new tool to for studying solid state physics\cite{ScRep}, such as the ferromagnetic-like behaviour discovered in this work as well as a new way of realising a magneto-mechanical system in the ultra-low loss regimes. It fits the new framework of magneto-mechanical systems that have recently emerged bringing together advantages of both spin and mechanical degrees of freedom. For example, such systems have been recently used for applications such as magnetometry\cite{Li:18} and optomechanical transduction\cite{Rudd2019}.
 
This work was supported by the Australian Research Council Grant No. CE170100009.

\section*{References}

\end{document}